\documentclass{amsart}
\usepackage{amssymb}
\usepackage{amscd}
\DeclareFontFamily{OMS}{cmsy}{%
\fontdimen16\font=3pt
\fontdimen17\font=3pt}
\makeatletter
\renewcommand{\subsection}{\@startsection{subsection}{2}{\z@}%
{\baselineskip}{0.5\baselineskip}{\bfseries}}
\makeatother
\def\dj{d\kern-.30em\raise1.25ex\vbox{\hrule width .3em height .03em}}
\def\Dj{D\rlap{\kern-.70em\raise0.75ex
\vbox{\hrule width .3em height .03em}}}
\def\dM{\mbox{$\smash{\vphantom{d}^{\raise0.4ex\hbox{$\scriptscriptstyle M$}}%
\!\!\!d}$}}
\def\bla#1{$(${\it #1\/{}}$)$}
\def\Q{\mathbb{Q}}
\def\N{\mathbb{N}}
\def\C{\mathbb{C}}
\def\Mes{\mathcal{M}}
\def\Si{\Sigma}
\def\Sj{\mathbb{S}}
\def\Z{\mathbb{Z}}

\def\im{\mathrm{im}}

\def\id{\mathrm{id}}

\renewcommand{\thepage}{\ifnum\value{page}=1 \else\arabic{page}\fi}
\newtheorem{thm}{Theorem}[section]
\newtheorem{pro}[thm]{Proposition}
\newtheorem{lem}[thm]{Lemma}
\theoremstyle{definition}
\newtheorem{defn}{Definition}
\newenvironment{pf}{\proof[\proofname]}{\endproof}
\newlength{\enviropost}
\newcommand{\be}{\begin{equation}}
\newcommand{\ee}{\end{equation}}
\newcommand{\ble}[1]{\begin{equation} \label{#1}}
\newcommand{\bae}{\begin{eqnarray}}
\newcommand{\eae}{\end{eqnarray}}
\newcommand{\fle}[2]%
{\vspace{1.5ex}
\be
\label{#1}
\mbox{%
\setlength{\fboxsep}{3ex}%
\framebox{$\dss #2 $}}
\ee} 
\newcommand{\nn}{\nonumber}
\newcommand{\ff}{\nn \\}
\newcommand{\fe}{& = &}

\newtheorem{state}{S$\! \!$}
\newtheorem{defin}{D$\! \!$}
\newtheorem{exatitle}{Example}
{\begin{exatitle} \label{#2} #1 \end{exatitle}}% 
{\hfill $\Box$ \vspace{\enviropost} \\}
{\vspace{3mm} \noindent {\bf Aside:} \small}% 
{\hfill \rule[-3mm]{0mm}{0mm}$\Diamond$\\}
{%
\vspace{3mm}  
\begin{center}
\begin{minipage}{.8\textwidth} 
\begin{state} 
\label{#1}%
}% 
{%
\end{state}
\end{minipage}
\end{center}
\vspace{3mm}
}
\newenvironment{definition}[1]%
{%
\vspace{3mm}  
\begin{center}
\begin{minipage}{.8\textwidth} 
\begin{defin} 
\label{#1}%
}% 
{%
\end{defin}
\end{minipage}
\end{center}
\vspace{3mm}
}
\newcommand{\dss}{\displaystyle}

\newcommand{\ip}[2]{\left\langle #1, #2\right\rangle}

\newcommand{\eg}{\hbox{\em e.g.{}}}

\newcommand{\ie}{\hbox{\em i.e.{}}}

\newcommand{\calA}{\mathcal{A}}

\newcommand{\calL}{\mathcal{L}}

\begin{document}
%%%%%%%%%%%%%%%%%%%%%%%%%%%%%%%%%%%%%%%%%%%%%%%%%%%%%%%%%%%%%%%%%%%%
%%%%%%%%%%%%%%%%%%%%%%%%%%%%%%%%%%%%%%%%%%%%%%%%%%%%%%%%%%%%%%%%%%%%
\title[generalized measures and Hopf algebras]{Higher Order
Measures,
Generalized\\
Quantum Mechanics and Hopf Algebras}
\thanks{\today}
\author{chryssomalis CHRYSSOMALAKOS \& micho {\Dj}UR{\Dj}EVICH} 
\address{Insituto de Ciencias Nucleares, 
UNAM, Area de la Investigacion Cientifica,
Circuito Exterior, Ciudad Universitaria, M\'exico DF, cp 04510,
MEXICO}
\email{ chryss@nuclecu.unam.mx}

\address{Instituto de Matematicas,
UNAM, Area de la Investigacion Cientifica,
Circuito Exterior, Ciudad Universitaria, M\'exico DF, cp 04510,
MEXICO}

\email{ micho@matem.unam.mx\newline
\indent\indent http://www.matem.unam.mx/{\~\/}micho}

\begin{abstract}
We study Sorkin's proposal of a generalization of
quantum mechanics and find that the theories proposed derive
their probabilities from $k$-th order polynomials in additive
measures, in the same way that quantum mechanics uses a
probability bilinear in the quantum amplitude and its complex
conjugate. Two complementary approaches are presented,
a $C^*$ and a Hopf-algebraic one, illuminating both algebraic
and geometric aspects of the problem. 

\end{abstract}
\maketitle

%%%%%%%%%%%%%%%%%%%%%%%%%%%%%%%%%%%%%%%%%%%%%%%%%%%%%%%%%%%%%%%%%%%%
%%%%%%%%%%%%%%%%%%%%%%%%%%%%%%%%%%%%%%%%%%%%%%%%%%%%%%%%%%%%%%%%%%%%
\section{Introduction}
\label{Intro}
%%%%%%%%%%%%%%%%%%%%%%%%%%%%%%%%%%%%%%%%%%%%%%%%%%%%%%%%%%%%%%%%%%%%
%%%%%%%%%%%%%%%%%%%%%%%%%%%%%%%%%%%%%%%%%%%%%%%%%%%%%%%%%%%%%%%%%%%%
In a series of papers~\cite{Sor:94,Sor:97}, Sorkin has put
forward a
view of quantum mechanics as a ``quantum measure theory".
His approach views the transition from classical to quantum
mechanics as a generalization of the additivity properties of the
classical measure function on a set of histories. 
This generalization has a
natural extension, producing a whole family of theories, indexed
by a positive integer $k$, each
defined by a particular ``sum rule" imposed on the measure
function. Our purpose in this paper is to show that the various
theories thus obtained can be characterized by the fact that the
corresponding measure is a polynomial of degree $k$ in primitive
(\ie, additive) functionals. 

In Sect.~2, the problem is approached from an algebraic
point of view. Sect.~3 complements the analysis in a
geometrical spirit, using the language of Hopf algebras.
Sect.~4, somewhat independent from the rest of the paper, 
sketches the relevance of these matters to theories
designed to overcome the obstacles to locality imposed by
Bell's inequalities. A concluding section suggests that
experiments should be done to establish the $k$ of nature
and points to formal interconnections with an already
existing concept of $k$-primitiveness in the literature. 

We start with a description of the two-slit interference
experiment, following closely the exposition
in~\cite{Sor:94,Sal:02}. Referring to the standard two-slit
setup,
we call $H$ the set of all electron histories (worldlines) leaving
the electron gun and arriving at the detector at specified time
instants (to avoid technicalities we consider $H$ to be discrete). 
We call $A$ ($B$) the subset of
$H$ consisting of all histories in which the electron passes
through slit $a$ ($b$) (we ignore the possibility of the
electron winding around both slits). Consider the four
possible ways of blocking the two slits and denote by 
$P_{ab}$, $P_a$, $P_b$ and $P_0=0$  the corresponding
probabilities of arrival at the detector,
the last one corresponding to both slits being
blocked off. The idea now is
to consider these probabilities as the values of a certain
measure function $\mu$ defined on the set of subsets of $H$, \eg,
$P_{a}=\mu(A)$. When mutually exclusive alternatives exist, as
when both slits are open, the union of the corresponding
(disjoint) subsets is to be taken, \eg, $P_{ab}=\mu(A \sqcup B)$
($\sqcup$ denotes disjoint union). Physical
theories are distinguished by their measures, for example, classical
mechanics uses a ``linear" measure $\mu_\text{cl}$,
satisfying the sum rule
\ble{sr2}
I^{\, \mu_\text{cl}}_2(A,B) \equiv \mu_\text{cl}(A \sqcup B)
-\mu_\text{cl}(A) - \mu_\text{cl}(B)=0
\, ,
\ee
and hence fails to account for any interference. Quantum
mechanics uses $\mu_\text{q}$, satisfying
$I^{\, \mu_\text{q}}_2(A,B)\neq 0$, as is well known. Sorkin's
observation was that in a three slit experiment (with eight
possibilities for blocking the slits), the probabilities predicted by
quantum mechanics {\em do} satisfy the sum rule
\bae
\label{sr3}
I^{\, \mu_\text{q}}_3(A,B,C) 
& \equiv & 
\mu_\text{q}(A \sqcup B \sqcup C)
-\mu_\text{q}(A \sqcup B) 
- \mu_\text{q}(A \sqcup C)
- \mu_\text{q}(B \sqcup C)
\ff
 & &
{}+\mu_\text{q}(A)
+\mu_\text{q}(B)
+\mu_\text{q}(C)
\ff
 \fe 0
\, ,
\eae
arguably a lesser known fact. It is easy to show that
$\mu_\text{cl}$ also satisfies~(\ref{sr3}), 
as a result of~(\ref{sr2}). There is an obvious generalization to
the $k$-slit experiment, involving the symmetric functional
$I^\mu_k$, given by
\bae
\label{Ikdef}
I^\mu_k (A_1, \ldots,A_k) 
& \equiv &
\mu(A_1 \sqcup \ldots \sqcup A_k)
\ff
 & &
{} -\sum_i 
\mu(A_1 \sqcup \ldots \sqcup \hat{A_i} \sqcup \ldots \sqcup
A_k)
\ff
 & &
{}+\sum_{i<j} 
\mu(A_1 \sqcup \ldots \sqcup \hat{A_i} \sqcup \ldots 
\sqcup \hat{A_j} \sqcup \ldots \sqcup
A_k)
\ff
 & & \ldots
\ff
 & &
{} +(-1)^{k+1} \sum_i \mu(A_i)
\, ,
\eae
where the hats denote omission and all $A_i$ are mutually
disjoint. These functionals satisfy the
recursion relation
\bae
\label{recIk}
I^\mu_{k+1}(A_0,A_1,\ldots,A_k)
\fe
I^\mu_k(A_0 \sqcup A_1,A_2,\ldots,A_k)
\ff
 & & 
{}-I^\mu_k(A_0,A_2,\ldots,A_k)
-I^\mu_k(A_1,A_2,\ldots,A_k)
\, ,
\eae
which implies that the sum rule $I^\mu_{k+1}=0$ follows from
$I^\mu_k=0$. One may now contemplate a family of theories,
indexed by a positive integer $k$, defined by the sum rule
$I^\mu_{k+1}=0$, with $I^\mu_k \neq 0$
for the corresponding measure. Classical mechanics is seen to be
a $k=1$ theory while quantum mechanics corresponds to $k=2$. 

The above formulas for $I^\mu_k$ need to be extended to the
general case, \ie, when the arguments are possibly overlapping
sets. For the $k=2$ case, Sorkin gives the following equivalent
forms
\bae
\label{k2over}
I^\mu_2
\fe
\mu(A \cup B) + \mu(A \cap B) - \mu(A \backslash B) 
- \mu(B \backslash A)
\ff
 \fe
\mu(A \bigtriangleup B) + \mu(A) + \mu(B) -2 \mu(A \backslash B)
-2 \mu(B \backslash A)
\, ,
\eae
derived by demanding bilinearity (the symbol $\backslash$ above
denotes set-theoretic difference while $\bigtriangleup$ denotes
symmetric difference).
%%%%%%%%%%%%%%%%%%%%%%%%%%%%%%%%%%%%%%%%%%%%%%%%%%%%%%%%%%%%%%%%%%%%
%%%%%%%%%%%%%%%%%%%%%%%%%%%%%%%%%%%%%%%%%%%%%%%%%%%%%%%%%%%%%%%%%%%%
\section{The Formalism of Non-linear Measures}
\label{TFNLM}
%%%%%%%%%%%%%%%%%%%%%%%%%%%%%%%%%%%%%%%%%%%%%%%%%%%%%%%%%%%%%%%%%%%%
%%%%%%%%%%%%%%%%%%%%%%%%%%%%%%%%%%%%%%%%%%%%%%%%%%%%%%%%%%%%%%%%%%%%

%%%%%%%%%%%%%%%%%%%%%%%%%%%%%%%%%%%%%%%%%%%%%%%%%%%%%%%%%%%%%%%%%%%%
\subsection{Preliminary considerations} 
\label{Pc}
%%%%%%%%%%%%%%%%%%%%%%%%%%%%%%%%%%%%%%%%%%%%%%%%%%%%%%%%%%%%%%%%%%%%
In the spirit of the functional theoretic
formulation of the classical measure theory, we are now going
to introduce an algebraic setup. The idea is to move from the
language of sets to the language of functions, replacing the
notion of measure by that of integral. 

Let us consider a unital $\Q$-algebra $A$. We shall deal with certain non-linear functionals 
$$ \mu\colon A\rightarrow \C $$
defined by a hierarchy of interesting algebraic relations.

For each $n\in\N$, let $\Mes_n(A)$ be the space of all maps $\mu$ satisfying 
\begin{equation}\label{measure-def}
\mu(a_1+\dots+a_{n+1})=\sum_S (-)^{n-|S|}\mu(\sum_{i\in S} a_i)
\end{equation}
where the $S\subset \{1,\dots, n+1\}$ runs over all subsets satisfying $1\leq |S| \leq n$. 

It is easy to see that each $\Mes_n(A)$ is an $A$-bimodule, in a natural manner. The additive
structure is trivial, while the left and right multiplications are given by
$$ (x\mu y)(a)=\mu (yax)\qquad x,y\in A. $$

Also, every $\Mes_n(A)$ allows multiplications by complex numbers (it is a complex 
vector space). Let us denote by $\Si_n(A)$ the space of multiadditive maps 
$$\Phi\colon \overbrace{A\times \dots \times A}^n \rightarrow \C$$ 
which are totally symmetric. The elements of $\Si_n(A)$ are naturally interpretable
as {\it homogeneous polynomials} of order $n$ over $A$. 
%%%%%%%%%%%%%%%%%%%%%%%%%%%%%%%%%%%%%%%%%%%%%%%%%%%%%%%%%%%%%%%%%%%%
\subsection{Quadratic measures}
\label{Qm}
%%%%%%%%%%%%%%%%%%%%%%%%%%%%%%%%%%%%%%%%%%%%%%%%%%%%%%%%%%%%%%%%%%%%
We shall first analyze a special case of {\it quadratic} measures (corresponding 
to $n=2$). As mentioned in the introduction, this completely covers probability 
aspects of standard quantum mechanics. Because of the importance of this 
special case, we shall present all calculations independently of the general setting, 
which will be discussed in the next subsection. 

Let us consider an arbitrary $\mu\in\Mes_2(A)$. The elements $\mu$ are characterized 
by the following identity
\begin{equation}\label{quadratic-def}
\mu(a+b+c)=\mu(a+b)+\mu(a+c)+\mu(b+c)-\mu(a)-\mu(b)-\mu(c), \quad \forall a,b,c\in A
\end{equation}

As the first elementary consequence, it is worth observing that 
$$ \mu(0)=0. $$

Furthermore, the group $\Z_2$ naturally acts on the space $\Mes_2(A)$. The action is
induced by right multiplication by $-1\in A$. It follows immediately that the space 
$\Mes_2(A)$ is naturally decomposed into a direct sum 
\begin{equation}\label{quadratic-dec}
\Mes_2(A) = \Mes_2^-(A)\oplus \Mes_2^+(A)
\end{equation}
of `even' and `odd' subspaces:
\begin{equation}
\begin{aligned}
\Mes_2^-(A) &= \Bigl\{ \mu \Bigm\vert \mu(a)=-\mu(-a)  \Bigr\} \\
\Mes_2^+(A) &= \Bigl\{ \mu \Bigm\vert \mu(a)=\mu(-a) \Bigr\}
\end{aligned}\qquad \forall a \in A. 
\end{equation}

Let us first analyze the odd part. As the following lemma shows, there is nothing
very exciting about $\Mes_2^-(A)$. 

\begin{lem} The space $\Mes_2^-(A)$ consists precisely of $\Q$-linear maps 
$\mu\colon A\rightarrow \C$. 
\end{lem}

\begin{pf}                                                
It is obvious that all $\Q$-linear $\mu$ belong to $\Mes_2^-(A)$. Let us observe that 
$\Q$-linearity is equivalent to {\it additivity}. Therefore what remains is to prove that
every $\mu\in\Mes_2^-(A)$ is additive. Indeed, replacing $c=-b$ in \eqref{quadratic-def}, 
and using the imparity assumption, we get
$$ \mu(a+b)+\mu(a-b)= 2\mu(a). $$
Replacing $a$ and $b$ in the above identity we obtain
$$ \mu(a+b)-\mu(a-b)= 2\mu(b). $$
Now summing the two equations we finally conclude
$$\mu(a+b)=\mu(a)+\mu(b)$$
which completes the proof. 
\end{pf}

The space $\Mes_2^+(A)$ possesses a much more interesting structure. For a given 
$\mu\in \Mes_2^+(A)$ let us define a map $\Phi\colon A\times A \rightarrow \C$ by 
\begin{equation}\label{quadratic-bili}
\Phi(a,b)=\frac{1}{4} \bigl( \mu(a+b)-\mu(a-b) \bigr). 
\end{equation}

It follows immediately that 
\begin{equation}
\Phi(a,b)=\Phi(b,a) 
\, ,
\end{equation}
\ie, the map $\Phi$ is symmetric. 

\begin{lem} \bla{i} The map $\Phi$ is $\Q$-bilinear. In other words
\begin{equation}
\Phi(\lambda a+b,c)=\lambda\Phi(a,c)+\Phi(b,c)\qquad \forall
a,b,c\in A,\quad \lambda\in \Q. 
\end{equation}

\bla{ii} We can reconstruct $\mu$ from $\Phi$ by
\begin{equation}\label{bili-quad}
\mu(x)= \Phi(x,x). 
\end{equation}
The correspondence $\mu\leftrightarrow \Phi$ is a natural isomorphism between 
the space $\Si_2(A)$ of symmetric bilinear functionals over $A$ and the space of 
even quadratic measures $\Mes_2^+(A)$. 
\end{lem}

\begin{pf}
Let us observe that $\Q$-bilinearity is equivalent to biadditivity. Using 
\eqref{quadratic-def} and performing elementary transformations we obtain
\begin{multline*}
\Phi(a, b+c) = \frac{1}{4} \bigl( \mu(a+b+c)-\mu(a-b-c) \bigr)
                 = \frac{1}{4} \bigl( \mu(a+b)+\mu(a+c)+\mu(b+c)\\
                 -\mu(a-b)-\mu(a-c)-\mu(b+c)\bigr)=\Phi(a,b)+\Phi(a,c) 
\, ,
\end{multline*}
which proves \bla{i}. Now using the bilinearity property of $\Phi$ we find 
$$ \Phi(x, x)= 4 \Phi(x/2, x/2) = \mu(x)-\mu(0)=\mu(x). $$

Finally, it is straightforward to see that every $\Phi\in \Si_2(A)$ gives rise, 
via \eqref{bili-quad}, to an even element $\mu\in\Mes_2(A)$. 
\end{pf}
%%%%%%%%%%%%%%%%%%%%%%%%%%%%%%%%%%%%%%%%%%%%%%%%%%%%%%%%%%%%%%%%%%%%
\subsection{Higher-order generalizations}
\label{Hog}
%%%%%%%%%%%%%%%%%%%%%%%%%%%%%%%%%%%%%%%%%%%%%%%%%%%%%%%%%%%%%%%%%%%%
In this subsection we shall generalize the previous analysis for arbitrary 
degrees $n\in\N$. Let us introduce, in an algebraic analogy with
\cite{Sor:94}, functionals 
\begin{equation}\label{i-fun}
I^\mu_k(a_1, \dots, a_k)=\sum_S (-)^{k-|S|}\mu(\sum_{i\in S} a_i), 
\end{equation}
where $k\geq 2$, the sumation is over all non-empty subsets 
$S\subseteq\{1, \dots, k\}$
and $\mu\colon A \rightarrow \C$ is an arbitrary map. By definition, all the maps $I_k$ are 
symmetric. 

\begin{lem}\label{i-mult} Let us assume that $\mu$ is arbitrary. We have
\begin{equation*}
I^\mu_{k+1}(b, c, a_2, \dots, a_k)=I^\mu_k(b+c, a_2, \dots, a_k)-I^\mu_k(b, a_2, \dots, a_k)
-I^\mu_k(c, a_2, \dots, a_k)
\end{equation*}
for each $a_i, b, c \in A$ and $k\geq 2$. 
\end{lem}

\begin{pf} This is matter of a straightforward combinatorial calculation, involving sums over 
different types of index subsets $S$: those that `contain' 
both $b$ and $c$, subsets containing only symbols $b$ or $c$, and those $S$ excluding
symbols $b$ and $c$. 
\end{pf}

It is easy to see that the following equivalences hold,
\begin{equation}\label{i-criterion}
\mu\in\Mes_n(A) \Leftrightarrow I^\mu_{n+1}=0  
\, ,
\qquad
\qquad
I^\mu_{n+1}=0 \Leftrightarrow I^\mu_{n} \in \Si_n(A)
\, .
\end{equation}

Taking into account the previous lemma, we conclude that $\mu\in\Mes_n(A)$ if
and only if the functional $I^\mu_n$ is multiadditive (and hence $\Q$-multilinear).  Hence, 
in this case we have $I^\mu_n\in \Si_n(A)$. Furthermore, we find 

\begin{equation}\label{i-inclusion}
\Mes_{n-1}(A)\subseteq \Mes_n(A), 
\end{equation}
in other words, $\Mes_k(A)$ form a monotonically increasing family of $A$-modules.

From now on, let us assume that $\mu\in \Mes_n(A)$ and define a map 
$\Phi\colon A^{\times n}\rightarrow \C$ by 
\begin{equation}
\Phi(a_1, \dots, a_n)=\frac{1}{2^nn!}\sum_z(-)^z\mu(z_1a_1+\dots+z_n a_n)
\end{equation}
where $z_i\in\{1,-1\}$ and $z=(z_1, \dots, z_n)$. 
           
\begin{lem} The following identity holds
\begin{equation}\label{i-phi}
I^\mu_n(a_1, \dots, a_n)=n!\Phi(a_1, \dots, a_n). 
\end{equation} 
\end{lem}

\begin{pf} A direct calculation gives 
\begin{multline*}
2^n I^\mu_n(a_1, \dots, a_n)=\sum_z (-)^z I^\mu_n (z_1a_1, \dots, z_n a_n)
=\sum_{z, S} (-)^{z+n-|S|} \mu\bigl(\sum_{i\in S} z_ia_i \bigr)\\
=\sum_z(-)^z\mu \bigl(\sum_{i=1}^n z_ia_i \bigr)=2^n n!\Phi(a_1, \dots, a_n) 
\end{multline*} 
and hence \eqref{i-phi} holds. We have applied the multilinearity property of $I_n$ 
in the above calculation. 
\end{pf}

Let us denote by $\Pi_n\colon \Mes_n(A)\rightarrow \Si_n(A)$ a linear map defined 
by $\Pi_n(\mu)=\Phi$. Using the previous lemma, and \eqref{i-inclusion} we find
\begin{equation}
\ker(\Pi_n)=\Mes_{n-1}(A). 
\end{equation}

The map $\Pi_n$ is really a projecton, and it admits a natural right section. 
Let us define $\iota_n\colon \Si_n(A)\rightarrow \Mes_n(A)$ by 
\begin{equation}
\mu(x)=\Phi(\overbrace{x, \dots, x}^n) \qquad \mu=\iota_n(\Phi). 
\end{equation}
Before going further, we have to verify that the image of $\iota_n$ is indeed within
the space $\Mes_n(A)$. A direct calculation gives
\begin{equation*}
\begin{split}
\sum_S(-)^{n-|S|}\mu(\sum_{i\in S} a_i)=\sum_S(-)^{n-|S|}\Phi(\sum_{i\in S} a_i, 
\dots ,\sum_{i\in S} a_i) \\
=\sum_S(-)^{n-|S|}\sum_\alpha\Phi(a_{i_1}, \dots, a_{i_n})=\sum_\alpha\Phi(a_{i_1}, \dots, a_{i_n})
=\mu(a_1+\dots+a_{n+1})
\end{split}
\end{equation*} 
where $\alpha=(i_1, \dots, i_n)$. The sumation is over subsets 
$S\subset \{1, \dots, n+1\}$ satisfying $1\leq |S| \leq n $. The last equality is obtained as 
follows. Let us focus on an index term 
$(i_1, \dots, i_n)$ having exactly $k$ different elements. The coefficient of this term 
is calculated by counting all the enveloping subsets $S$, with the 
corresponding signs.  We arrive at 
$$
\sum_{l=k}^n(-)^{n-l}\binom{n+1-k}{l-k} =(-)^{n-k}\sum_{l=0}^{n-k}(-)^l\binom{n+1-k}{l}
=-(-)^{n-k}(-)^{n-k+1}=1. 
$$
Hence, $\im(\iota_n)\subseteq\Mes_n(A)$. It is easy to see that 
\begin{equation}
\Pi_n\iota_n(\Phi)=\Phi\qquad\forall\Phi\in\Si_n(A). 
\end{equation}
Indeed, for $\mu=\iota_n(\Phi)$ we have 
\begin{equation*}
\begin{split}
\Pi_n(\mu)(a_1, \dots, a_n)=\frac{1}{2^nn!}\sum_z (-)^z\mu(z_1a_1+\dots+z_na_n)\\=
\frac{1}{2^nn!}\sum_z(-)^z\Phi(\sum_iz_ia_i,\dots,\sum_iz_ia_i)\\
=\frac{1}{2^nn!}\sum_z(-)^z\sum_\alpha z_{i_1}\dots z_{i_n}\Phi(a_{i_1}, \dots, a_{i_n})
=\Phi(a_1, \dots, a_n). 
\end{split}
\end{equation*}
The last equality is obtained by observing that only
multi-indexes $\alpha=(i_1, \dots, i_n)$ 
that are {\it permutations} count, as other terms would cancel each other. The factor 
$2^nn!$ emerges as we sum over $\Z_2^n\times S_n$. 

Summarizing our considerations we can now formulate 
\begin{pro} For every $n\geq 2$, there is a natural split short exact sequence
\begin{equation}
0\xrightarrow{} \Mes_{n-1}(A) \hookrightarrow \Mes_n(A) 
\xrightarrow{\Pi_n} \Si_n(A)\xrightarrow{} 0 \qquad \iota_n\colon \Si_n(A)\rightarrow \Mes_n(A)
\end{equation} 
which allows us to introduce a canonical decomposition
\begin{equation}
\Mes_n(A)\leftrightarrow \Mes_{n-1}(A)\oplus \Si_n(A). 
\end{equation}

The elements of $\Mes_n(A)$ are nothing but polynomial functions of order (less than
or equal to) $n$. In terms of the above identification, the elements of $\Si_n(A)$ correspond
to homogeneous polynomials of order $n$. \qed

\end{pro}

%%%%%%%%%%%%%%%%%%%%%%%%%%%%%%%%%%%%%%%%%%%%%%%%%%%%%%%%%%%%%%%%%%%%
\section{Hopf Algebras and Generalized Measures}
\label{HAGM}
%%%%%%%%%%%%%%%%%%%%%%%%%%%%%%%%%%%%%%%%%%%%%%%%%%%%%%%%%%%%%%%%%%%%
%%%%%%%%%%%%%%%%%%%%%%%%%%%%%%%%%%%%%%%%%%%%%%%%%%%%%%%%%%%%%%%%%%%%
%%%%%%%%%%%%%%%%%%%%%%%%%%%%%%%%%%%%%%%%%%%%%%%%%%%%%%%%%%%%%%%%%%%%
\subsection{Hopf algebras}
\label{Ha}
%%%%%%%%%%%%%%%%%%%%%%%%%%%%%%%%%%%%%%%%%%%%%%%%%%%%%%%%%%%%%%%%%%%%
We give here a few basic definitions about Hopf algebras and some 
intuitive comments concerning their content. We keep the
discussion informal, our basic aim being to point out the
relevance of Hopf algebraic concepts to the problem at hand.

Restricted to the
cocommutative case (we explain the term below), which is the one
of interest here, the axioms for a Hopf 
algebra are just dual to those for a group. The duality is the
one between points of the group manifold $G$ and functions on the
manifold and is formally expressed via an inner product,
\ble{ipdef}
\ip{\cdot}{\cdot} : \quad \calA \otimes G \rightarrow
\mathbb{C},
\qquad \qquad
f \otimes g \rightarrow \ip{f}{g}
\, \equiv \, f(g)
\, ,
\ee
extended by linearity to the group algebra.
$\calA \equiv \text{Fun}(G)$ is the (commutative) algebra of 
complex valued 
functions on $G$ while the last equation above simply states that
the duality mentioned is by pointwise evaluation. The definition
of a group involves the notions of a product $m:\, G \otimes G
\rightarrow G$, an identity $e \in G$ and an
inverse, which dualize, via the above inner
product, to the notion of a {\em coproduct} $\Delta$,
\ble{copnot}
\Delta: \, \calA \rightarrow \calA \otimes \calA
\, ,
\qquad \qquad
f \mapsto \Delta(f) 
\, \equiv \, 
\sum_i f_{(1)}^i \otimes f_{(2)}^i 
\, \equiv \, 
f_{(1)} \otimes f_{(2)}
\, ,
\ee
a {\em counit} $\epsilon$,
\ble{epsnot}
\epsilon : \, \calA \rightarrow \mathbb{C}
\, ,
\qquad \qquad
f \mapsto \epsilon(f)
\, ,
\ee
and a {\em coinverse} or {\em antipode} $S$,
\ble{Snot}
S: \, \calA \rightarrow \calA 
\, ,
\qquad \qquad
f \mapsto S(f)
\, ,
\ee
respectively. The defining relations are
\bae
\label{Hadef}
\ip{f}{m(g \otimes g')} 
\, = \, 
\ip{f}{g g'} 
& \equiv & 
\ip{f_{(1)} \otimes f_{(2)}}{g \otimes g'}
\, = \,
\ip{f_{(1)}}{g} \ip{f_{(2)}}{g'}
\ff
\epsilon(f)
& \equiv & 
\ip{f}{e}
\ff
\ip{S(f)}{g}
& \equiv & 
\ip{f}{g^{-1}}
\, ,
\eae
\ie, the Hopf algebraic operations are the adjoints, with respect
to the above inner product, of those of a group%
\footnote{%
The above, although suitable for our purposes, is not the
standard definition of a Hopf algebra. The latter can be
consulted in, \eg, Ref.~\cite{Swe:69}.%
}%
. When the group is abelian (as in our case), the coproduct 
satisfies $f_{(1)}
\otimes f_{(2)} = f_{(2)} \otimes f_{(1)}$ --- in this case the
Hopf algebra is called {\em cocommutative}. Notice that
\ble{hom}
\Delta(fh)=\Delta(f)\Delta(h)
\, ,
\qquad \, \,
\epsilon(fh)=\epsilon(f) \epsilon(h)
\, ,
\qquad \, \, 
S(fh)=S(h)S(f)
\, ,
\ee
where the product in $\calA \otimes \calA$ is defined by 
$(f \otimes h)(f' \otimes h') = ff' \otimes hh'$ (the primes
distinguish functions here, they do not denote differentiation). 
Dual to the associativity of the  group product is the  {\em
coassociativity} of the coproduct,
\ble{coassoc}
(\Delta \otimes \id) \circ \Delta  = (\id \otimes \Delta)
\circ \Delta
\, .
\ee
Then the notation $\Delta^k$ is unambiguous, since it does not
matter to which tensor factor are the successive $\Delta$'s
applied to --- the resulting function of $k+1$ arguments will be
denoted by $f_{(1)} \otimes \ldots \otimes f_{(k+1)}$ and it is
invariant, in the cocommutative case, under exchange of any two
tensor factors. Notice finally that dual to the property of the
unit $ge= eg
= g$ is the property of the counit 
\ble{couprop}
\epsilon(f_{(1)}) f_{(2)} = f_{(1)} \epsilon(f_{(2)}) =f
\, .
\ee
%%%%%%%%%%%%%%%%%%%%%%%%%%%%%%%%%%%%%%%%%%%%%%%%%%%%%%%%%%%%%%%%%%%%
\subsection{Coderivatives}
\label{Coder}
%%%%%%%%%%%%%%%%%%%%%%%%%%%%%%%%%%%%%%%%%%%%%%%%%%%%%%%%%%%%%%%%%%%%
One way of looking at the coproduct of a function is as
an indefinite translation. Indeed, defining the
right translation $R_g$ on the group by $R_g(g')=g'g$, its
pullback on functions $R_g^*(f) \equiv f_g$ is given by
$f_g(g')=f(g'g)=f_{(1)}(g') f_{(2)}(g)$, which shows that
$f_{(1)}(\cdot') f_{(2)}(g)$ is the right-translated $f$ (by $g$),
while $f_{(1)}(\cdot') f_{(2)}(\cdot)$, a function of two
arguments,  is the indefinitely translated $f$, with the second
argument defining the translation and the first evaluating the
translated function (one obtains a left version of the above
exchanging the two tensor factors of the coproduct). With this in
mind, one recognizes the operator $\calL: \, \calA \mapsto
\calA \otimes \calA$,
defined by
\ble{Ldef}
\calL f = \Delta(f) - f \otimes 1
\, ,
\ee
as a (dualized) {\em indefinite discrete derivative} or {\em
coderivative} for short,
\bae
\label{idLd}
(\calL f)(g',\, g)
\fe
\ip{f_{(1)}\otimes f_{(2)} - f \otimes 1}{g' \otimes g} 
\ff
 \fe
f(g'g)-f(g')
\, .
\eae
When $g$ is close to the identity, $g=e+X+\ldots$, with
$X$ in the Lie algebra of the group, $(\calL f)(\cdot',g)$ is
(proportional to) the derivative of $f$ along the left
invariant vector field corresponding to $X$. One may define
higher order coderivatives $\calL^k f$, with the
understanding that the successive applications of $\calL$ are to
be taken at the leftmost tensor factor, 
\ble{Lkdef}
\calL^{k} f \equiv (\calL \otimes \id) \circ \calL^{k-1} f
\, ,
\qquad 
k=2,3,\ldots
\, \, ,
\ee
so that, for example,
\bae
\label{L2ex}
\calL^2 f 
& \equiv &
(\calL \otimes \id) \circ \calL f 
\ff
 \fe
(\calL \otimes \id)  (f_{(1)} \otimes f_{(2)} - f \otimes 1)
\ff
 \fe
f_{(1)} \otimes f_{(2)} \otimes f_{(3)}
- f_{(1)} \otimes 1 \otimes f_{(2)} 
\ff
 & &
 {}-f_{(1)} \otimes f_{(2)} \otimes 1
+ f \otimes 1 \otimes 1
\, .
\eae 
Of particular interest to us will be the evaluation of the above
$k$-th order coderivative at the identity of the
group, $(\calL^k f)(e,\cdot, \ldots)\equiv (\calL^k f)(e)$, \eg,
\bae
\label{L2ex2}
(\calL f)(e)
\fe
f - \epsilon(f) 1
\ff
(\calL^2 f)(e)
\fe
f_{(1)} \otimes f_{(2)} - f \otimes 1 -1 \otimes f + \epsilon(f)
1 \otimes 1
\, ,
\eae
where~(\ref{couprop}) was used. We are now ready to introduce the
basic notion of {\em $k$-primitiveness}
%%%%%%%%%%%%%%%%%%%%%%%%%%
\begin{definition}{kprimdef}
A function $f$ will be called {\em $k$-primitive} if all its 
coderivatives at the identity $(\calL^r f)(e)$, $r > k$ are
equal to zero, while $(\calL^k f)(e)$ is not. 
\end{definition}
%%%%%%%%%%%%%%%%%%%%%%%%%%
%%%%%%%%%%%%%%%%%%%%%%%%%%%%%%%%%%%%%%%%%%%%%%%%%%%%%%%%%%%%%%%%%%%%
\subsection{Generalized quantum mechanics and $k$-primitiveness}
\label{Gqmkp}
%%%%%%%%%%%%%%%%%%%%%%%%%%%%%%%%%%%%%%%%%%%%%%%%%%%%%%%%%%%%%%%%%%%%
%%%%%%%%%%%%%%%%%
\subsubsection{Group structure on the set of histories}
\label{gssh}
%%%%%%%%%%%%%%%%%
Consider the set of histories $H$ associated to some given
experiment, taken as a discrete set for simplicity.
For a subset $A$ of $H$, let $\chi_A$ be the
{\em characteristic function} of $A$, defined by $\chi_A(x)=1$ if
$x\in A$, $\chi_A(x)=0$ if $x \in H \backslash A$. It is clear
that one may deal with the subsets of $H$ in terms of their
characteristic functions, as we do in the following. Denote by $G$
the set of all linear combinations of characteristic functions of 
subsets of $H$, \ie, a typical element $g$ of $G$ is of the form
$g= \lambda_1 \chi_{A_1} + \lambda_2 \chi_{A_2} + \ldots$, where the
$A_i$ are subsets of $H$ and $\lambda_i \in \mathbb{C}$. 
We may turn $G$ into an abelian group
defining the group law by addition. Then for the identity $e$ we
have $e= \chi_\emptyset=0$ and the inverse of $g$ is $-g$. 

Just like in the introduction, a physical theory derives its
probabilities from a measure function $\mu$, defined now on
$G$, \eg, $P_a=\mu(\chi_A)$ in the two-slit experiment. When
mutually exclusive alternatives exist, the {\em sum of the
characteristic functions of the corresponding subsets} is to be
taken. Notice that, in terms of the subsets themselves,
this corresponds to disjoint union, in
accordance with the operation used in~\cite{Sor:94},
\cite{Sal:02}. The important point is that simply by extending
this definition (\ie, addition of the characteristic functions) to 
non-disjoint subsets we recover the rather
complicated interference term~(\ref{k2over}) and its
generalizations, as we now show. Indeed, consider a quadratic
functional $\mu^2$, with $\mu$ additive, evaluated on two overlaping 
subsets $A$ and $B$  --- the resulting interference term is
\bae
\label{recint}
I^{\mu^2}_2 
\fe
\mu(\chi_A + \chi_B)^2
-
\mu(\chi_A)^2 - \mu(\chi_B)^2
\ff
 \fe
 2 \mu(\chi_A) \mu(\chi_B)
\ff
 \fe
2 \big( 
\mu(\chi_{A \backslash B}) \mu(\chi_{B \backslash A})
+ \mu(\chi_{A \backslash B}) \mu(\chi_{A \cap B})
\ff
 & &
{}+ \mu(\chi_{A \cap B}) \mu(\chi_{B \backslash A})
+ \mu(\chi_{A \cap B})^2
\big)
\, ,
\eae
where, in the last step, we wrote $\chi_A=\chi_{A \backslash B} +
\chi_{A \cap B}$ and similarly for $\chi_B$.
On the other hand, the first, for example, of~(\ref{k2over})
becomes
\ble{calcint}
I^{\mu^2}_2 
=
\mu(\chi_{A \cup B})^2 + \mu(\chi_{A \cap B})^2 -\mu(\chi_{A
\backslash B})^2 - \mu(\chi_{B \backslash A})^2
\, .
\ee
Substituting 
$\chi_{A \cup B}=\chi_{A \backslash B} + \chi_{B \backslash A} +
\chi_{A \cap B}$ and expanding one recovers the right hand side
of~(\ref{recint}).
%%%%%%%%%%%%%%%%%
\subsubsection{$k$-primitive functions on $G$}
\label{kpfG}
%%%%%%%%%%%%%%%%%
We focus now on the commutative and cocommutative Hopf algebra 
$\calA \equiv \text{Fun}(G)$. Among its elements are the quantum
measures $\mu$ we have been considering so far.
The fact that 
$\mu(\emptyset)=0$ translates, in the Hopf algebraic
language of this section, into the statement that the counit of all 
measures vanishes, $\mu(e)=\epsilon(\mu)=0$. The linearity of the
classical measure, Eq.~(\ref{sr2}), becomes here the statement
that $\mu_\text{cl}(\chi_A + \chi_B)= \mu_\text{cl}(\chi_A) +
\mu_\text{cl}(\chi_B)$, which is easily seen to dualize to
\ble{muclH}
0 = (\calL^2 \mu)(e) = \mu_{\text{cl} \, (1)} 
\otimes \mu_{\text{cl} \, (2)} 
- \mu_\text{cl} \otimes 1 - 1 \otimes \mu_\text{cl} +
\epsilon(\mu_\text{cl}) 1 \otimes 1
\, ,
\ee
the last term being zero. Hence, according to (D1),
$\mu_\text{cl}$ is a 1-primitive element of $\calA$. 
More generally, we have the following
\begin{lem}
The symmetric functionals $I^\mu_k$, defined in Eq.~(\ref{Ikdef}),
coincide with the $k$-th order coderivatives $(\calL^k \mu)(e)$
of Eq.~(\ref{Lkdef}).
\end{lem}
We omit the straightforward inductive proof. We may now state the
main result of this section
\begin{pro}
In the algebra $\calA$ of functions on $G$, 
every $k$-primitive element 
is a $k$-th degree polynomial in 1-primitive elements.
\end{pro}
\begin{pf}
$G$, being abelian, admits 1-primitive coordinates, \eg, the
normal ones, which we call $\xi_i$. Any element of $\calA$, in
particular a $k$-primitive measure $\mu$, 
is a function of the $\xi_i$, $\mu=\mu(\xi_i)$. From the
vanishing of $(\calL^{k+1} \mu)(e)$, with $(\calL^k \mu)(e) \neq
0$, one may infer, by evaluating on arguments infinitesimaly
close to the identity of $G$, that $(X_{i_1} \ldots
X_{i_{k+1}})(\mu)(e)=0$, for all $X_i$ in the Lie algebra of $G$,
while $(X_{i_1} \ldots X_{i_{k}})(\mu)(e) \neq 0$ for at least
one index set.
Given that $X_i (\mu) = \frac{\partial \mu}{\partial \xi_j} X_i
(\xi_j)$, one infers that $\frac{\partial^{k+1} \mu}{\partial
\xi_{j_1} \ldots \partial \xi_{j_{k+1}}}(e)=0$, for all $j_i$,
while at least one $k$-th order partial derivative is non-zero at
the identity. The proposition is then proved by repeated
integration.
\end{pf}
The same conclusion can be reached by establishing that $\calA$ is 
a cocommutative graded connected Hopf algebra and hence, by
applying the Milnor-Moore theorem~\cite{Mil.Moo:65},  
isomorphic to the universal enveloping algebra of its subalgebra
of 1-primitive elements. We point out that in~\cite{Sal:02}, it
has been observed that if $\mu$ is the $k$-th power of a ``linear" 
functional then $I^\mu_r =0$, for $r>k$.  
%%%%%%%%%%%%%%%%%%%%%%%%%%%%%%%%%%%%%%%%%%%%%%%%%%%%%%%%%%%%%%%%%%%%
\section{A $C^*$-algebraic formulation}
\label{ACsf}
%%%%%%%%%%%%%%%%%%%%%%%%%%%%%%%%%%%%%%%%%%%%%%%%%%%%%%%%%%%%%%%%%%%%
We are going to touch upon some interesting issues related to a 
$C^*$-algebraic formulation of the 
algebraic setup of Sect.{} 2. A special emphasis will 
be given to possible relationships 
between the introduced formalism, the theory of contextulal hidden 
variables \cite{Mi5}, and a
corresponding non-Kolmogorovian probability framework as a way 
of overcoming obstacles to locality, 
given by Bell's inequalities. In order to keep this section 
reasonably short, we will only sketch
basic ideas, and leave detailed presentations with proofs for 
another article (in such a way we will have 
more published papers, which is {\it good}). 
 
We will assume here that $A$ is a $C^*$-algebra. 
By definition \cite{BR} this means that $A$ is 
a Banach algebra, equiped with a *-structure (antilinear and 
antimultiplicative involution), so that 
\begin{equation*}
|aa^*| = |a|^2 \qquad\quad \forall a\in A. 
\end{equation*}
 A remarkable property 
of $C^*$-algebras is that the norm is uniquely fixed by the 
algebra structure. In other words, for a given 
*-algebra $A$, there at most one $C^*$-algebraic norm. 
In such a way $C^*$-algebras form a 
full subcategory of complex *-algebras. 

The algebra $A$ is called unital if there is a (necessarily 
unique) unit element $1\in A$.  We will deal with 
unital algebras only. 

From the point of view of our 
considerations, we can think of $A$ as consisting of {\it physical observables}.\footnote{More precisely, 
hermitian elements of $A$ are viewed as physical observables.} Two special cases are the most 
interesting here: 

\begin{itemize}
\item Classical case---$A$ is a commutative algebra, generated by certain functions on the 
system's phase space $\Gamma$. For example, we can assume $A=\mathbb{M}(\Gamma)$. That is, $A$ is 
the algebra of (classes of) essentially bounded measurable functions on $\Gamma$. 

\item Quantum case---$A$ is a non-commutative algebra, generated by operators acting in the Hilbert
state space $H$. For example, we can assume $A=\mathbb{B}(H)$. In other words, $A$ is the algebra of 
all bounded operators acting in $H$. 
\end{itemize}

However, all our considerations apply to general $C^*$-algebras. Let us begin by recalling the concept
of a {\it state}. This is any functional $\rho\colon A\rightarrow \mathbb{C}$ satisfying 
\begin{gather*}
\rho(a^* a)\geq 0 \qquad \forall a\in A\\
\rho(1)=1. 
\end{gather*}
In other words, a state is a positive and normalized functional on $A$. It is easy to see that the set of all 
states on $A$ is convex, and compact in the *-weak topology
of the dual space $A^*$. According to the Krein-Millman 
theorem, $\mathbb{S}(A)$ is the closure of the convex hull of its extremal elements. These extremal elements are
called {\it pure states}. 

The theory of states generalizes the classical probability theory, to the level of non-commutative (quantum) 
spaces. Indeed, if $A$ is commutative then, according to the classical Gelfand-Naimark theory, we have 
a natural identification
\begin{equation*}
A \leftrightarrow C(X)
\end{equation*}
where $X$ is a compact topological space---the spectrum of $A$ (the set of all characters $\kappa\colon 
A\rightarrow \mathbb{C}$, equipped with the *-weak topology of the dual space $A^*$). In this 
commutative case, states on $A$ correspond, according to the 
classical Riesz representation theorem, to 
probability measures on $X$. The correspondence is given by the Lebesgue integral. 

Taking into account the considerations of Sect.~\ref{TFNLM}, 
it is natural to formulate 
\begin{defn} A generalized, order-$n$, state on $A$ is a map 
$\rho\in \Si_n(A)$ satisfying
\begin{gather*}
\rho(a^*a)\geq 0 \qquad \forall a\in A\\
\rho(1)=1. 
\end{gather*}
\end{defn}
Let us denote by $\Sj_n(A)$ the set of such order-$n$ states on $A$. It is easy to see that $\Sj_n(A)$ is 
convex, and can be equipped with a natural *-weak topology, converting it into a compact topological 
space. Applying the Krein-Millman theorem, it follows that $\Sj_n(A)$ is the closure of the convex hull of its 
extremal elements. 

\begin{defn}
The extremal elements of $\Sj_n(A)$ are called pure (order-$n$) states on the algebra $A$. 
\end{defn}

Let us assume that $A$ is generated by its projectors (hermitian idempotents $p=p^*=p^2$). Such elements 
correspond to elementary yes/no situations, and can be viewed as the simplest possible physical 
observables. We can also identify projectors with {\it events}. In the classical case projectors correspond 
to the appropriate subsets of the phase space. 

So, given a projector $p\in A$ and a higher-order state $\rho\colon A\rightarrow \mathbb{C}$, we want 
to interpret the number $\rho(p)\in [0,1]$ as the {\it probability of the event $p$ in the state $\rho$}, just as in the 
standard case (otherwise, it would not be much of a sense to formulate the above definitions). 

However, higher-order states contain an additional obstacle for such an interpretation. Let us consider 
two events $p,q\in A$ that are realizable simultaneously (within the same experimental context, this 
implies that $pq=qp$). Let us also assume that $p$ and $q$ are mutually exclusive. This means that 
$pq=qp=0$ so we have orthogonal projectors. If our higher-order state $\rho$ represents something really 
meaningfull, then we must have
\begin{equation}\label{pq-lin}
\rho(p+q)=\rho(p)+\rho(q).
\end{equation}
The above condition is automatically fulfilled for standard states (due to linearity). For all higher-order states, 
the condition is actually a condition for $p$ and $q$.  In particular, if we put $q=1-p$ we get a 
non-trivial algebraic condition on a single event $p$. In other words, {\it not all events are allowed}. Of course, it 
might happen that for a given $\rho$ there are no non-trivial projectors $p$ satisfying the consistency condition. 
In this case, the state $\rho$ is basically useless, from the point of view of the statistical interpretation of 
its values on projectors. On the other hand, the states that always satisfy the consistency condition (for every 
orthogonal events $p$ and $q$) are, in all non-perverted scenarios, just the standard linear states. This follows 
from the generalized Gleason theorem by Maeda \cite{Maeda}. 

Therefore, in order for a higher-order state $\rho$ to be {\it reasonable}, it should have sufficiently many 
`good' projectors $p$. This motivates our next definition.

\begin{defn} Let us consider a higher-order state $\rho\in\Sj_n(A)$. A projector $p\in A$ is called 
$\rho$-compatible if 
\begin{equation}
\rho(p)+\rho(1-p)=1. 
\end{equation} 
The state $\rho$ is called $A$-compatible if the set of all $\rho$-compatible projectors generates the whole 
$C^*$-algebra $A$. 
Finally, for a given $A$-compatible state $\rho$, a unital $C^*$-subalgebra $B\subseteq A$ is called 
$\rho$-compatible, if \eqref{pq-lin} holds for all mutually orthogonal projectors from $B$. 
\end{defn}

Let us assume that $\rho$ is an arbitrary $A$-compatible higher-order state. Then it gives rise to a nice short 
exact sequence of $C^*$-algebras: 

\begin{equation}\label{ext-A}
0\xrightarrow{} \mathcal{K} \hookrightarrow \widehat{A}
\xrightarrow{\pi} A\xrightarrow{} 0
\end{equation}
Here $\widehat{A}$ is the free $C^*$ algebra generated by all $\rho$-compatible subalgebras $B$ of $A$. The
map $\pi\colon\widehat{A}\rightarrow A$ is the natural projection homomorphism and $\mathcal{K}=\ker(\pi)$. 

In a special case when $A$ is commutative, the above exact sequence is very similar to a class of contextual 
subquantum extensions considered in \cite{Mi5}. Indeed, we can write $A=C(\Omega)$ where $\Omega$ is 
the spectrum of $A$ (interpreted here as the sub-quantum space
of the system) and our extension becomes: 
\begin{equation}\label{ext-subq}
0\xrightarrow{} \mathrm{com}(A) \hookrightarrow \widehat{A}
\xrightarrow{\pi} C(\Omega)\xrightarrow{} 0
\end{equation}
The $\rho$-compatible subalgebras $B$ correspond to allowed {\it measurement contexts} in $\Omega$. We see that the 
probability theory on $\Omega$ is a non-Kolmogorovian one,
in the case of higher-order states $\rho$: the additivity 
of the measure holds only within the measurement contexts. The non-commutative algebra $\widehat{A}$ 
corresponds to the {\it full subquantum algebra}. The kernel of $\pi$ is simply the commutant of $A$. 
Such models overcome obstacles to locality given by Bell's inequalities, because they are based on the appropriate 
non-Kolmogorovian statistics. The composite systems are simply described by taking tensor products of the 
introduced extensions. 

\section{Conclusions and Final Remarks}
\label{CaFR}
%%%%%%%%%%%%%%%%%%%%%%%%%%%%%%%%%%%%%%%%%%%%%%%%%%%%%%%%%%%%%%%%%%%%
%%%%%%%%%%%%%%%%%%%%%%%%%%%%%%%%%%%%%%%%%%%%%%%%%%%%%%%%%%%%%%%%%%%%
We have studied Sorkin's hierarchy of generalizations of quantum
mechanics and found that the $k$-th order generalized measures
are necessarily $k$-th degree polynomials in 1-primitive
functionals, in the same sense that standard quantum mechanics
derives its probabilities from a bilinear expression in
a 1-primitive (\ie, additive) quantum amplitude and its
(also 1-primitive) complex conjugate. The question
of how is positivity to be attained in a, for example, cubic
theory is still open. On the other hand, one may envisage a
$k=4$ theory as a small cuartic correction to the standard
quantum mechanical probability, showing up as a small
deviation of $I^\mu_4$ from zero in a four-slit experiment. 
What we find remarkable is the 
immediacy
with which the sum rules $I^\mu_k=0$ connect to a simple $k$-slit
experiment, a point that might be worth bringing to the attention
of our experimental colleagues. 

On a more formal level, a very important subject is the study of 
interrelations between states and representations of
$C^*$-algebras generated by physical observables. According
to the GNS construction \cite{BR}, there is a natural one-to-one 
correspondence
\begin{equation*} 
\left\{ \mbox{Standard states $\rho$ on $A$}\right\} 
\Leftrightarrow \left\{ \mbox{Equivalence classes 
of cyclic representations of $A$}\right\}
\end{equation*}
In terms of this correspondence, pure states translate into 
irreducible representations. The generalization of 
the GNS construction for the higher-order states introduced in
this article is a subject for further research. 

It is worth mentioning that the extensions of commutative
$C^*$-algebras of Sect.~\ref{ACsf} 
by non-commutative ones
play a central role in algebraic $K$-theory and non-commu\-tative 
geometry \cite{BDF, WOlsen}. For example, non-commutative
extensions similar to~(\ref{ext-subq}) can be used to build a
$K$-homology theory for metrizable compact topological spaces
$\Omega$.

We finish by pointing out that a concept of $k$-primitiveness has
appeared recently in the study of the Hopf algebra structure in
the process of renormalization in quantum field
theory (see Ref.~\cite{Bro.Kre:00}). A second, esssentially
equivalent, definition was given and the concept was further analyzed
in~\cite{Chr.Que.Ros.Ver:01}. In those works it refers to the 
much more
complicated case of the Hopf algebra of rooted trees but these
earlier definitions
can be shown to be identical with the one presented
here, although, we feel, the latter clarifies the underlying 
geometrical
content. We plan on further elucidating these interconnections in
an upcoming publication.  
%%%%%%%%%%%%%%%%%%%%%%%%%%%%%%%%%%%%%%%%%%%%%%%%%%%%%%%%%%%%%%%%%%%%
%\bibliographystyle{plain}
%\bibliography{strings}
%\begin{thebibliography}{20} 
%\bibitem{KN} Kobayashi S, Nomizu K:{\it Foundations of Differential Geometry},
%Interscience Publishers, New York, London (1963) 
%\bibitem{Sor} Sorkin R: {\it Quantum Mechanics as Quantum Measure Theory},
%Preprint gr-qc/9401003 (1994)
%\end{thebibliography}
%%%%%%%%%%%%%%%%%%%%%%%%%%%%%%%%%%%%%%%%%%%%%%%%%%%%%%%%%%%%%%%%%%%%

\end{document}